# Decoding User Concerns in AI Health Chatbots: An Exploration of Security and Privacy in App Reviews

Muhammad Hassan
University of Illinois at Urbana-Champaign
mhassa42@illinois.edu

Abdullah Ghani
and Muhammad Fareed Zaffar
Lahore University of Management Sciences
{25100155,fareed.zaffar}@lums.edu.pk

Masooda Bashir
University of Illinois at Urbana Champaign
mnb@illinois.edu

*Abstract*—AI-powered health chatbot applications are increasingly utilized for personalized healthcare services, yet they pose significant challenges related to user data security and privacy. This study evaluates the effectiveness of automated methods, specifically BART and Gemini GenAI, in identifying security-privacy-related (SPR) concerns within these applications' user reviews, benchmarking their performance against manual qualitative analysis. Our results indicate that while Gemini's performance in SPR classification is comparable to manual labeling, both automated methods have limitations, including the misclassification of unrelated issues. Qualitative analysis revealed critical user concerns, such as data collection practices, data misuse, and insufficient transparency and consent mechanisms. This research enhances the understanding of the relationship between user trust, privacy, and emerging mobile AI health chatbot technologies, offering actionable insights for improving security and privacy practices in AI-driven health chatbots. Although exploratory, our findings highlight the necessity for rigorous audits and transparent communication strategies, providing valuable guidance for app developers and vendors in addressing user security and privacy concerns.

## I. INTRODUCTION & BACKGROUND

AI-powered health chatbots are transforming the healthcare landscape by providing accessible, efficient, and scalable solutions for users seeking medical advice, symptom checking, and mental health support [7], [11], [25] These applications leverage advanced natural language processing (NLP) techniques to simulate human-like interactions, offering users immediate assistance without requiring direct contact with healthcare professionals. As their adoption grows, AI health chatbots have the potential to bridge gaps in healthcare accessibility, particularly in underserved regions or during crises such as the COVID-19 pandemic [25].

However, the increasing reliance on these applications raises critical privacy and security concerns. Health chatbots often process sensitive personal information, including medical histories, symptoms, and demographic data [2]. Mishandling this data—whether through inadequate security measures, unauthorized sharing with third parties, or insufficient transparency—can lead to severe consequences such as data breaches, identity theft, or violations of regulatory frameworks like HIPAA (Health Insurance Portability and Accountability Act) and GDPR (General Data Protection Regulation) [26], [9], [13], [19]. For instance, recent studies on mobile health applications have identified critical vulnerabilities, exposing millions of users' private health data and emphasizing the urgent need for enhanced privacy and security measures in this rapidly evolving domain [17].

While prior research has extensively examined privacy and security issues in mobile health applications [16], [6], limited attention has been paid to the unique challenges posed by AI-driven health chatbots [22]. These systems introduce novel risks due to their reliance on sensitive personal data and algorithmic decision-making processes that lack transparency [22]. This gap is significant because these chatbots introduce unique risks due to their reliance on machine learning models that process and store sensitive user data. Moreover, user feedback—often expressed through app reviews—provides valuable insights into real-world privacy and security concerns that may not surface in controlled studies or developer disclosures [12], [22]. Understanding how users articulate these concerns is essential for improving trust in AI-driven technologies. It also ensures compliance with regulatory standards.

In this context, sentiment analysis tools such as BART and other natural language processing (NLP) models have been widely adopted to classify user feedback and identify thematic concerns in app reviews [23]. These tools are highly effective for analyzing sentiment and detecting user priorities [8]. However, there is limited research comparing automated methods like BART with generative AI models (e.g., Gemini) or manual labeling for classifying security-privacy-related (SPR) reviews. Furthermore, while thematic analysis is well-established in qualitative research [4], its application to SPR concerns in AI-driven health chatbots has yet to be fully explored. This paper addresses this gap by focusing on how users express privacy and security concerns in app reviews of AI-powered health chatbots. Specifically, we make the following key contributions:

1) **Comparative Analysis of Labeling Methods:** We evaluate three distinct methods for identifying security-privacy-related (SPR) concerns in user reviews: two automated approaches using BART (Bidirectional and Auto-Regressive Transformers) Large MNLI and Gemini GenAI models, and manual labeling by human reviewers. By comparing these methods, we aim to determine their effectiveness in accurately classifying SPR reviews at scale.
2) **Analysis of User Concerns:** Using app reviews as a lens, we study the types of privacy and security concerns users raise about AI-powered health chatbot applications. This analysis provides actionable insights into user priorities and highlights gaps in existing practices.

These contributions are significant for several reasons. First, understanding user concerns directly from app reviews allows us to capture real-world issues that may be overlooked by surveys or developer-centric evaluations. App reviews often reflect users' lived experiences with these technologies, making them an invaluable resource for identifying pain points related to privacy and security. Second, our comparative analysis of labeling methods provides a foundation for future research on scalable approaches to analyzing large datasets of user feedback. By evaluating both machine learning models and human coding approaches, we offer practical guidance on balancing efficiency with accuracy when studying privacy/security concerns at scale.

In summary, this paper seeks to advance the understanding of user privacy and security concerns in AI-powered health chatbot applications by leveraging app reviews as a rich source of information. Our findings aim to inform developers, policymakers, and researchers about how to design more secure and trustworthy systems that align with user expectations and regulatory requirements.

## II. USER REVIEW LABELING

In this section, we describe our methodology for comparing the effectiveness of three approaches—BART, Gemini, and manual labeling—for identifying and classifying security-privacy-related (SPR) user reviews in AI-powered health chatbot applications. This study focuses on Android applications due to their widespread adoption and significant user base among mobile consumers [21].

**App Selection:** We selected applications from the Google Play Store that provide AI-powered health chatbot services. Search queries such as "Health Chatbot," "Medical Chatbot," "Health GPT," and "AI Doctor" were used to identify relevant applications. From the combined search results, we filtered free applications with over 10,000 downloads and at least one user review. Table I provides an overview of the selected applications.

**Review Collection:** After selecting the applications, we extracted user reviews using the Google Play Scraper framework [20]. For this exploratory study, we randomly sampled 1,000 reviews from applications with more than 1,000 total reviews. For apps with fewer than 1,000 reviews, all available reviews were collected. This process resulted in a dataset comprising 5,469 user reviews.

### A. Labeling User Reviews

In order to compare the effectiveness of three methods—BART, Gemini, and manual labeling—for classifying user reviews of AI-powered health chatbot applications into security-privacy-related reviews and non-SPR reviews. The goal is to identify user-reported privacy and security concerns in these applications. Prior research has shown that SPR reviews can be identified using various strategies, such as analyzing negative or low-rated reviews [15]. However, recent studies have demonstrated that SPR reviews may also be present in highly rated reviews [18], necessitating a more comprehensive sampling strategy. Therefore, in this study, we sampled reviews irrespective of their ratings to ensure a broader analysis.

We employed automated and manual labeling approaches to classify reviews as SPR or non-SPR. Automated methods leverage machine learning models to classify reviews efficiently but may face challenges in understanding nuanced user concerns. Manual labeling offers higher accuracy but is time-intensive and subject to human biases. To address these limitations, we conducted a comparative evaluation of these methods as part of this study.

| App Name | Installs | Score | Developer | Genre |
|---|---|---|---|---|
| Ada – check your health | 10,315,071 | 4.6125 | Ada Health | Medical |
| Wysa: Anxiety, therapy chatbot | 4,250,928 | 4.611894 | Touchkin | Health & Fitness |
| Youper - CBT Therapy Chatbot | 1,401,603 | 3.9500272 | Youper, Inc | Health & Fitness |
| Sintelly: CBT Therapy Chatbot | 1,146,601 | 4.3512545 | Sintelly | Health & Fitness |
| Symptomate – Symptom checker | 502,816 | 4.29 | Infermedica | Health & Fitness |
| AI Dermatologist: Skin Scanner | 449,931 | 4.5348835 | Acina | Health & Fitness |
| HealthPal - AI Health Advisor | 426,425 | 4.58 | Bigwell Lab | Tools |
| Dr.Oracle AI Medical Assistant | 11,599 | 4.5882354 | TheDeep | Medical |

TABLE I: Overview of selected AI Health Chatbots applications with their respective installs, scores, developers, and genres.

*1) Automated Labeling*: For automated labeling, we utilized two state-of-the-art models: BART (Bidirectional and Auto-Regressive Transformers), Large MNLI, and Gemini GenAI. These models were chosen for their advanced natural language processing capabilities and their ability to classify textual data into meaningful categories. BART, with its strong performance in multi-genre natural language inference, excels at discerning subtle distinctions in sentiment and intent, making it particularly suitable for identifying security and privacy concerns. In juxtaposition, Gemini GenAI represents a cutting-edge approach that leverages generative AI techniques, allowing for a more dynamic understanding of user reviews and the ability to generate contextually relevant labels.

**BART:** BART is a transformer-based model designed for natural language understanding and generation tasks [27]. It combines bidirectional context modeling (like BERT) with autoregressive generation (like GPT), making it highly effective for tasks such as text classification. In this study, we applied the zero-shot classification pipeline using the `bart-large-mnli` model on the dataset of 5,469 reviews. The model was run twice to generate two sets of labels: sentiment and SPR concern.

- Sentiment labels: Positive, Negative, Neutral
- Concern labels: "Privacy-Security Concern," "Other Concern," "No Concern"

Additionally, we collected confidence scores for each label predicted by BART to assess the reliability of its classifications.

**Gemini:** Gemini is a generative AI model designed for advanced natural language understanding across diverse contexts [24]. Its ability to discern subtle differences in sentiment and intent makes it well-suited for classifying user reviews into SPR and non-SPR categories. For this study, we used Gemini GenAI Model 1.5 Flash to label the dataset for sentiment and SPR concerns. The model was prompted to provide structured responses that included both the sentiment label and the SPR label for each review, along with confidence scores for its predictions. The specific prompt used for this task is provided in Appendix A.

*2) Manual Labeling*: To establish a ground truth dataset for comparison with automated methods, we manually labeled approximately 20% of the dataset (1,100 out of 5,469 reviews). These 1,100 reviews were randomly sampled from the curated dataset used for automated labeling by BART and Gemini. The manual labeling process involved two reviewers independently categorizing 1,100 reviews for sentiment and SPR concerns. Inter-rater reliability was measured using Cohen's Kappa coefficient to ensure consistency.

- Sentiment labels: Cohen's Kappa = 0.71
- SPR concern labels: Cohen's Kappa = 0.66

These values indicate substantial agreement between the reviewers, and the remaining minor disagreements were mutually resolved to obtain a high-quality gold dataset.

As a result of manual labeling, the 1,100 reviews had the following distributions: for the Sentiment labels, 926 reviews were classified as Positive, 133 as Negative, and 43 as Neutral; for the Concern labels, 7 reviews were categorized as Privacy-Security Concern, 144 as Other Concern, and 949 as No Concern.

Interestingly, one of the seven Privacy-Security concerns had a positive sentiment, while none were neutral, highlighting the diversity in sentiment even within critical concerns.

*B. Results*

In this section, we compare the performance of BART and Gemini in assessing review sentiment and identifying SPR reviews, using our manually labeled dataset as the ground truth. For both BART and Gemini, the label with the highest score assigned by the model was selected as its verdict.

**Metrics:**

To holistically evaluate the performance of both models, we employ standard metrics: accuracy, precision, recall, and F1-score. Moreover, macro-averaging is used to account for the imbalanced nature of the dataset for metrics except for accuracy.

**Sentiment Classification:**

Figure 1 illustrates the performance of models for sentiment classification. Gemini outperforms BART across all metrics by significant margins, achieving 96% accuracy in predicting the sentiment of reviews compared to 81% for BART. Additionally, Gemini demonstrates greater stability across all sentiment categories, achieving a macro-f1 score of 83%, which is 30% greater than BART.

According to Figure 4, a closer analysis suggests that identifying reviews with a neutral sentiment comes at a challenge; BART achieves a subpar 9% accuracy at classifying neutral reviews. On the other hand, both models demonstrate solid performance in identifying negative reviews, achieving 88% accuracy with Gemini and 90% with BART. Given the predominance of negative sentiment among SPR reviews, this suggests that sentiment classification can effectively aid in their identification.

**Concern Classification:**

The performance of both models in concern classification can be observed in Figure 5. Unlike sentiment analysis, BART achieves a subpar accuracy of 28%, performing worse than random chance for a three-class classification task. In contrast, Gemini achieves a much higher accuracy of 89%. However, Gemini's performance is limited in correctly identifying Privacy/Security concerns, successfully detecting only 4 out of the 7 such concerns in the labeled dataset. In contrast, BART accurately identifies all 7 SPR reviews but does so with a very low precision (3.18%). As shown in Figure 5, BART tends to associate a concern in the majority of cases, unlike Gemini, which adopts a much more conservative approach. While using BART minimizes the risk of missing critical SPR reviews, the approach becomes impractical and unscalable when applied to larger datasets.

## III. USER SPR CONCERNS

In this section, we analyze the security and privacy-related (SPR) concerns identified in user reviews classified as SPR reviews. These reviews provide insights into the challenges users face regarding the data security and privacy behaviors of AI-powered health chatbot applications. Given the preliminary nature of this study, these findings should not be considered exhaustive, as ongoing research may uncover additional SP concerns.

### A. Manual Qualitative Analysis of Security and Privacy Concerns

The identification and analysis of SP concerns were conducted in a two-step process. Initially, one researcher independently reviewed the manually labeled SPR reviews and assigned security and privacy-related themes to the reviews. Subsequently, these themes were refined and validated through discussions with a second researcher. The findings are summarized in the following key themes:

*1) Data Collection and Surveillance Concerns:* A significant portion of user concerns revolved around the data collection practices of these applications, particularly regarding the collection and monitoring of user information. Users often questioned the relevance and necessity of certain data requests, expressing skepticism about the legitimacy of these practices. For instance, one user remarked:

> *"Why does this app have to access my photo? Does not make sense at all."*

This comment reflects apprehension about the potential safety of media files and a lack of clarity about the app's data requirements. Furthermore, some users expressed dissatisfaction with the functionality of the applications, especially when they failed to operate as expected after collecting user data:

> *"Don't work. Probably a scam to collect data."*

> *"It takes you a lot of time and effort to sign in, they ask a lot of questions, and at the end, you won't understand the report."*

Another user highlighted concerns about text message monitoring:

> *"It's fake. You don't get anything except as being installed onto your phone; it will also monitor all your text messages."*

Such reviews indicate a perception of intrusive monitoring and an urgent need for applications to articulate how user data is collected, processed, and protected clearly.

*2) Data Misuse and Sharing Concerns:* Respecting users' rights to understand how their data is used or shared is a fundamental principle of privacy[3]. However, many reviews revealed concerns about potential misuse or sharing of user data with third parties, often without adequate explanation. For example, a user stated:

> *"In the description, they claim they don't sell your data, but when you actually install the app, it turns out they do sell your data. Edit 23rd Oct 2024: The vendor has replied and claims your data is not sold, but this is not the truth. The app sends data to third parties based on so-called legitimate interest, and that's strictly illegal in the EU."*

This comment reflects a strong awareness of privacy laws among users, but also highlights the inadequacy of the app's communication about its compliance with privacy regulations. Even when data practices align with legal standards, unclear or ambiguous explanations can erode user trust.

*3) Transparency and Consent Mechanisms:* AI-powered health chatbots handle sensitive health and medical data, necessitating robust transparency and effective consent mechanisms. Yet, user reviews often pointed to inadequacies in these areas. Concerns were raised about data being used to train systems without explicit consent:

> *The accuracy of the diagnosis is 40–50%, that is, errors are quite frequent. It looks like the system is just learning and has little data. The idea is good, but trying to collect data and learn from users' money is bad.*

Similarly, the absence of effective consent mechanisms was a recurring issue:

> *Is there an option to create an account so it can remember what choices I have opted into It's annoying having to agree to terms and conditions, and give basic health information, every time I use the app.*

These reviews highlight the need for improved user-centric designs that prioritize transparency and seamless consent processes.

### B. Summary of Findings

The reviews analyzed reveal significant user concerns regarding data security and privacy practices in AI-powered health chatbots. These applications, by their very nature, handle sensitive health and medical information, necessitating adherence to stringent information security and data privacy principles. User feedback underscores the importance of enhanced transparency, robust consent mechanisms, and clear communication about data practices to foster trust and confidence in these systems.

## IV. DISCUSSION & FUTURE WORKS

In this exploratory study, we compared multiple methods for classifying security and privacy-related (SPR) reviews from a dataset of user feedback on AI-powered health chatbot applications. Specifically, we utilized the BART model, Gemini (a generative AI-based method), and manual labeling. Manual labeling was used as the ground truth to evaluate the performance of automated methods. Our findings indicate that Gemini performed comparably to manual labeling in terms of accuracy, demonstrating its potential as a scalable alternative for analyzing SPR reviews. However, nuances in review content revealed limitations in its reliability. Additionally, we investigated the specific security and privacy concerns expressed by users in the SPR dataset. Users frequently highlighted discomfort with intrusive data collection practices, apprehension about data sharing with third parties without explicit consent, and dissatisfaction with unclear or inadequate consent mechanisms[28], [1].

User reviews provide rich and meaningful insights into their concerns and issues. However, it is important to note that some of the concerns raised by users may not always be valid or justified. For example, users occasionally expressed apprehensions based on misunderstandings or incomplete knowledge of the applications' operations [10].

We observed that during the classification of SPR reviews, Gemini occasionally misclassified reviews unrelated to security and privacy. For instance, reviews mentioning loneliness or unrelated personal experiences were inaccurately categorized as SPR, potentially impacting the reliability of GenAI-based classification. To address this, we propose incorporating a well-defined definition of security-privacy concerns as part of the prompt when querying GenAI models [10]. However, this approach should be applied cautiously to avoid skewing results based on the specificity of the provided definition.

Our sentiment analysis revealed that SPR reviews are not exclusively negative. In fact, some reviews with positive sentiment also highlighted significant SP concerns [14]. This finding underscores the nuanced nature of user feedback, where users may appreciate certain aspects of an application while simultaneously raising serious privacy or security concerns.

### A. Implications for Practice

The emerging nature of AI-powered health chatbot technologies and their increasing adoption demand rigorous audits and continuous improvement in data safety and privacy practices. Users' concerns regarding data collection, sharing, and transparency highlight the need for strong regulatory compliance and enhanced design practices that prioritize user trust. The results of this study indicate that automated methods, Gemini GenAI in particular, can be effectively used to analyze SPR reviews. Additionally, these methods can support further manual qualitative analysis of SP concerns, enabling researchers and practitioners to extract actionable insights.

### B. Limitations and Future Directions

We acknowledge that this work is exploratory in nature, and its findings are not fully generalizable across all AI-powered health chatbots at this stage. The dataset used in this study was limited and focused on a subset of popular AI-powered health chatbot applications and reviews. Additionally, the manual labeling used as ground truth, while robust, may introduce subjective bias inherent to human reviewers. To improve the reliability and scope of our findings, future work will incorporate broader datasets and a wider variety of app reviews.

We plan to refine our GenAI prompts further and explore hybrid human-AI classification systems to reduce misclassification errors and enhance model interpretability [5]. Feedback from the presentation of this work will guide improvements, including expanding the dataset to encompass a more diverse set of applications and reviews.

In conclusion, our study demonstrates that SPR reviews in AI-powered health chatbots can be effectively studied using automated methods and qualitative analysis. However, these findings emphasize the need for ongoing efforts to ensure user privacy and data security in this rapidly evolving technological landscape.

## V. CONCLUSION

This study addressed two pivotal research questions: **(RQ1)** How effective are automated methods compared to manual analysis in identifying security-privacy-related (SPR) concerns in user reviews of AI-powered health chatbots? **(RQ2)** What specific SPR concerns do users raise about these applications?

To investigate the RQ1, we evaluated the performance of two automated approaches, BART and Gemini GenAI, against manual labeling. While Gemini demonstrated accuracy comparable to manual labeling for SPR classification, both methods showed limitations, such as misclassifying irrelevant reviews and failing to capture nuanced concerns. These results highlight the scalability and efficiency of automated methods but also underscore the need for refinement—such as integrating clearer prompts or employing hybrid human-AI approaches—to achieve the depth and contextual understanding of human analysis.

For the RQ2, manual qualitative analysis of manually labeled reviews revealed three primary SPR concerns. Users were apprehensive about data collection and surveillance, particularly regarding access to information perceived as unnecessary. Data misuse and unauthorized sharing with third parties, including potential breaches of privacy regulations, were also significant concerns. Additionally, users criticized the lack of transparency and inadequate consent mechanisms, emphasizing the need for clear communication of data handling practices and enhanced user controls.

The findings highlight the critical importance of embedding robust privacy and security measures into the design of AI-powered health chatbots. Developers must prioritize transparency, ensure compliance with privacy regulations, and embrace user-centered design principles to build trust in these sensitive applications. Ongoing future work focuses on refining automated SPR classification methods, expanding apps and reviews datasets, and exploring hybrid systems that leverage both human judgment and AI capabilities, aiming to advance and provide a robust framework for understanding and addressing users' security and privacy concerns in AI-driven health chatbot applications.

APPENDIX

```
1  You are an assistant specialized in analyzing app
       reviews. Your task is to classify reviews
       according to the following criteria, allowing
       for one or more labels as needed:
2
3  1. **Sentiment Analysis**:
4     - Classify the sentiment using one or more of the
          following labels: 'Positive', 'Negative',
          'Neutral'.
5     - Provide a confidence score (0-100) for each
          label included.
6     - Include a brief explanation supporting your
          sentiment classifications.
7
8  2. **Concern Identification**:
9     - Identify if the review raises concerns and
          label it with one or more of the following:
          'Privacy/Security Concern', 'Other Concern',
          'No Concern'.
10    - Attach a confidence score (0-100) for each
          label provided.
11    - Offer a brief explanation justifying your
          classifications.
12
13 Your response must strictly follow this JSON schema:
14 ```json
15 {
16   "sentiment_labels_confidence_tuple": [["label",
          confidence], ...],
17   "sentiment_explanation": "string",
     "concern_labels_confidence_tuple": [["label",
          confidence], ...],
     "concern_explanation": "string"
}
```

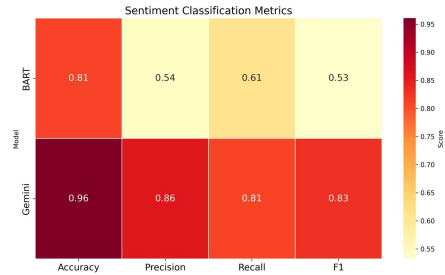

Fig. 1: Sentiment Classification

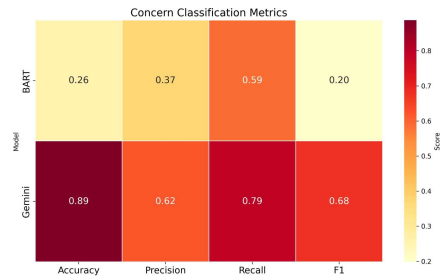

Fig. 2: Concern Classification

Fig. 3: Performance of models for (4) Sentiment Classification and (5) Concern Classification.

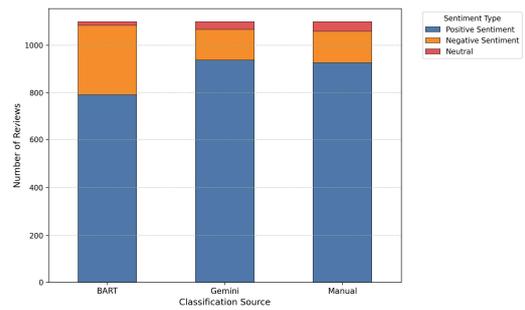

Fig. 4: Sentiment Distribution

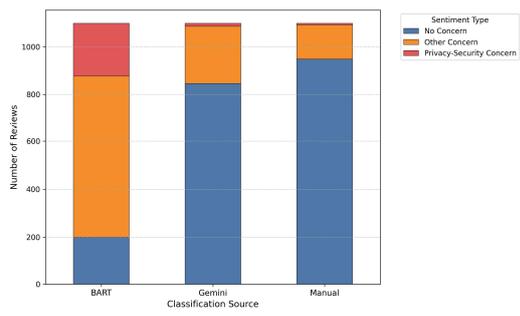

Fig. 5: Concern Distribution